\documentclass[a4,11pt]{article}
\usepackage{graphicx} 
\usepackage{amsmath}
\usepackage{amsfonts}
\usepackage{latexsym}
\usepackage{amssymb}
\makeatletter
 
  \@addtoreset{equation}{section}
 \makeatother

\begin{document}
\title{Collateralized CDS and Default Dependence~\footnote{
This research is supported by CARF (Center for Advanced Research in Finance) and 
the global COE program ``The research and training center for new development in mathematics.''
All the contents expressed in this research are solely those of the authors and do not represent any views or 
opinions of any institutions. 
The authors are not responsible or liable in any manner for any losses and/or damages caused by the use of any contents in this research.
}\\
{\it \Large -Implications for the Central Clearing-}
}

\author{Masaaki Fujii\footnote{Graduate School of Economics, The University of Tokyo},
Akihiko Takahashi\footnote{Graduate School of Economics, The University of Tokyo}
}
\date{
First version: April 11, 2011\\ 
}
\maketitle



\newtheorem{definition}{Definition}
\newtheorem{assumption}{$[$ A}
\newtheorem{condition}{$[$ C}
\newtheorem{lemma}{Lemma}
\newtheorem{proposition}{Proposition}
\newtheorem{theorem}{Theorem}
\newtheorem{remark}{Remark}
\newtheorem{example}{Example}
\newtheorem{corollary}{Corollary}
\def\n{{\bf n}}
\def\A{{\bf A}}
\def\B{{\bf B}}
\def\C{{\bf C}}
\def\D{{\bf D}}
\def\E{{\bf E}}
\def\F{{\bf F}}
\def\G{{\bf G}}
\def\H{{\bf H}}
\def\I{{\bf I}}
\def\J{{\bf J}}
\def\K{{\bf K}}
\def\L{{\bf L}}
\def\M{{\bf M}}
\def\N{{\bf N}}
\def\O{{\bf O}}
\def\P{{\bf P}}
\def\Q{{\bf Q}}
\def\R{{\bf R}}
\def\S{{\bf S}}
\def\T{{\bf T}}
\def\U{{\bf U}}
\def\V{{\bf V}}
\def\W{{\bf W}}
\def\X{{\bf X}}
\def\Y{{\bf Y}}
\def\Z{{\bf Z}}
\def\cala{{\cal A}}
\def\calb{{\cal B}}
\def\calc{{\cal C}}
\def\cald{{\cal D}}
\def\cale{{\cal E}}
\def\calf{{\cal F}}
\def\calg{{\cal G}}
\def\calh{{\cal H}}
\def\cali{{\cal I}}
\def\calj{{\cal J}}
\def\calk{{\cal K}}
\def\call{{\cal L}}
\def\calm{{\cal M}}
\def\caln{{\cal N}}
\def\calo{{\cal O}}
\def\calp{{\cal P}}
\def\calq{{\cal Q}}
\def\calr{{\cal R}}
\def\cals{{\cal S}}
\def\calt{{\cal T}}
\def\calu{{\cal U}}
\def\calv{{\cal V}}
\def\calw{{\cal W}}
\def\calx{{\cal X}}
\def\caly{{\cal Y}}
\def\calz{{\cal Z}}
%
\def\sskip{\hspace{0.5cm}}
\def\simleq{ \raisebox{-.7ex}{\em $\stackrel{{\textstyle <}}{\sim}$} }
\def\leqsim{ \raisebox{-.7ex}{\em $\stackrel{{\textstyle <}}{\sim}$} }
\def\ep{\epsilon}
\def\half{\frac{1}{2}}
\def\iku{\rightarrow}
\def\Iku{\Rightarrow}
\def\ikup{\rightarrow^{p}}
\def\inclusion{\hookrightarrow}
\def\cadlag{c\`adl\`ag\ }
\def\up{\uparrow}
\def\down{\downarrow}
\def\doti{\Leftrightarrow}
\def\douti{\Leftrightarrow}
\def\dochi{\Leftrightarrow}
\def\douchi{\Leftrightarrow}%
\def\yy{\\ && \nonumber \\}
\def\y{\vspace*{3mm}\\}
\def\nn{\nonumber}
\def\be{\begin{equation}}
\def\ee{\end{equation}}
\def\bea{\begin{eqnarray}}
\def\eea{\end{eqnarray}}
\def\beas{\begin{eqnarray*}}
\def\eeas{\end{eqnarray*}}
%
\def\hd{\hat{D}}
\def\hv{\hat{V}}
\def\hsd{{\hat{d}}}
\def\hx{\hat{X}}
\def\hsx{\hat{x}}
\def\bsx{\bar{x}}
\def\bsd{{\bar{d}}}
\def\bx{\bar{X}}
\def\ba{\bar{A}}
\def\bb{\bar{B}}
\def\bc{\bar{C}}
\def\bv{\bar{V}}
\def\balpha{\bar{\alpha}}
\def\bbalpha{\bar{\bar{\alpha}}}
\def\combi{\l(\begin{array}{c}\alpha\\ \beta \end{array}\r)}
\def\f{^{(1)}}
\def\s{^{(2)}}
\def\ss{^{(2)*}}
\def\l{\left}
\def\r{\right}
\def\a{\alpha}
\def\b{\beta}
\def\L{\Lambda}

\def\E{{\bf E}}
\def\P{{\bf P}}
\def\Q{{\bf Q}}
\def\R{{\bf R}}

\def\calf{{\cal F}}
\def\calp{{\cal P}}
\def\calq{{\cal Q}}

\def\ep{\epsilon}

\def\yy{\\ && \nonumber \\}
\def\y{\vspace*{3mm}\\}
\def\nn{\nonumber}
\def\be{\begin{equation}}
\def\ee{\end{equation}}
\def\bea{\begin{eqnarray}}
\def\eea{\end{eqnarray}}
\def\beas{\begin{eqnarray*}}
\def\eeas{\end{eqnarray*}}
\def\l{\left}
\def\r{\right}
\vspace{10mm}

\begin{abstract}
In this paper, we have studied the pricing of a continuously collateralized CDS. We have made use of 
the "survival measure" to derive the pricing formula in a straightforward way.
As a result, we have found that
there exists irremovable trace of the counter party as well as the investor in the price of CDS 
through their default dependence {\it even under the perfect collateralization}, although the hazard rates of the 
two parties are totally absent from the pricing formula.
As an important implication, we have also studied the situation where the investor enters an offsetting back-to-back 
trade with another counter party. We have provided simple numerical examples to 
demonstrate the change of a fair CDS premium according to the strength of default dependence 
among the relevant names, and then discussed its possible implications for the risk management of the central counter parties.

\end{abstract}
\vspace{17mm}

\newpage
\section{Introduction}
The recent financial crisis, exemplified by the collapse of Lehman Brothers, 
is the major driver of the increased use of collateral agreements based on
CSA (credit support annex published by ISDA), which is now 
almost a market standard among financial institutions~\cite{ISDA}.
Coupled with the explosion of various basis spreads, such as Libor-OIS  and 
the cross currency swap (CCS) basis, the effects of collateralization on the derivative pricing,
particularly from the view point of funding costs, have become important research topics.
Johannes \& Sundaresan (2007) \cite{collateralized_swap} have first emphasized the cost of 
collateral using swap rates in U.S. market. In a series of works Fujii \& Takahashi (2009, 2010)~\cite{multiple_curves, dynamic_basis, collateral_termstructures}, we have developed a framework of interest-rate modeling in the presence of collateralization 
and multiple currencies. We have also pointed out the importance of choice of collateral currency and 
the embedded cheapest-to-deliver option in collateral agreements~\cite{collateral_choice}.
Piterbarg (2010)~\cite{Piterbarg} discussed the general issue of option pricing under collateralization.

This financial crisis has also brought about serious research activity on the 
credit derivatives and counter party default risk,  and large amount of research papers 
have been published since the crisis. The regulators have also been working hard 
to establish the new rules for the counter party risk management,  and also for the migration toward
the CCPs (central counter parties) particularly in the CDS (credit default swap) market.
The excellent reviews and collections of recent works are available in the books 
edited by Lipton \& Rennie~\cite{Lipton} and Bielecki, Brigo \& Patras~\cite{Bielecki}, for example.
However, the effects of collateralization on credit derivatives remain still largely unclear.
It seems partly because that the idea of collateral cost appeared only recently, 
and also because the detailed collateral modeling is very complicated, due to the existence of 
settlement lag, threshold, and minimum transfer amount, e.t.c..

In this paper, based on the market development toward more stringent collateral management requiring 
a daily (or even intra-day) margin call, we have studied the pricing of CDS 
under the assumption of continuous collateralization. This is expected to be particularly relevant 
for the CCPs dealing with CDS and 
other credit linked products, for which the assumption of continuous collateralization 
seems to be a reasonable proxy of the reality.
Although we have studied the similar issues for the standard fixed income derivatives in the 
previous work~\cite{asymmetric_collateral}, the result cannot be directly applied to the credit derivatives
since the behavior of hazard rates generally violates the so called "no-jump" condition
(e.g. Collin-Dufresne, Goldstein \& Hugonnier (2004)~\cite{collin}) 
if there exists non-trivial default dependence among the relevant parties.
In this work, we 
apply the technique introduced by Sch\"{o}nbucher (2000)~\cite{Schonbucher} and adopted 
later by 
\cite{collin} in order to eliminate 
the necessity of this condition.

As a result, we have obtained a simple pricing formula for the collateralized CDS.
We will see, under the perfect collateralization, that the CDS price does not depend on the 
counter party hazard rates at all as expected. However, very interestingly, there 
remains irremovable trace of the two counter parties through the default dependence.
This gives rise to a very difficult question about the appropriate pricing method 
for CCPs. A CCP acts as a buyer as well as a seller of a given CDS at the same time,
by entering a back-to-back trade between the two financial firms. However, the result tells 
us that the mark-to-market values of the two offsetting CDSs are not equal in general even at the time 
of inception, if the CCP adopts the same price for the two firms. 
Although the detailed work will be left in a separate paper, we think that the result has
important implications for the proper operations of CCPs for credit derivatives.

\section{Fundamental Pricing Formula}
\subsection{Setup}
\label{setup}
We consider a filtered probability space $(\Omega, \calf, \mathbb{F},Q)$, where 
$Q$ is a spot martingale measure, and $\mathbb{F}=\{\calf_t~;~ t\geq 0\}$ is a sub-$\sigma$-algebra
of $\calf$ satisfying the usual conditions.  We denote the set of relevant firms $\calc=\{0,1,2, ..., n\}$ 
and introduce a strictly positive random variable $\tau^i$ in the 
probability space as the default time of each party $i\in \calc$. 
We define the default indicator process of each party as $H_t^i=\bold{1}_{\{\tau^i\leq t\}}$
and denote by $\mathbb{H}^i$ the filtration generated by this process.
We assume that we are given a background filtration $\mathbb{G}$ containing other information 
except defaults and write $\mathbb{F}=\mathbb{G}\vee \mathbb{H}^0\vee \mathbb{H}^1\vee\cdots\vee\mathbb{H}^n$.
Thus, it is clear that $\tau^i$ is an $\mathbb{H}^i$ as well as $\mathbb{F}$ stopping time.
We assume the existence of non-negative hazard rate process $h^i$ where 
\be
M_t^i=H_t^i-\int_0^t h_s^i \bold{1}_{\{\tau^i>s\}}ds, ~\quad t\geq 0
\ee
is an $(Q,\mathbb{F})$-martingale. We also assume that there is no simultaneous default
for simplicity.

For collateralization, we assume the same setup adopted in \cite{asymmetric_collateral} 
and repeat it here once again for 
convenience: {\it Consider a trade between the party $1$ and $2$.
If the party $i~(\in\{1,2\})$ has a negative mark-to-market value, it has to post the cash 
collateral~\footnote{According to the ISDA survey~\cite{ISDA}, more than $80\%$ of collateral 
being used is cash. If there is a liquid repo or security-lending market, we may also carry out 
similar formulation with proper adjustments of its funding cost.} to the 
counter party $j~(\neq i)$, where the coverage ratio of the exposure is denoted by $\delta_t^i\in\mathbb{R}_{+}$.
We assume the margin call and settlement occur instantly.
Party $j$ is then a collateral receiver and has to pay collateral rate $c_t^{i}$ on 
the posted amount of collateral, which is $\delta_t^{i}\times (|$mark-to-market$|)$, to the party $i$. 
This is done continuously until the end of the contract.
Following the market conventions, we set the collateral rate $c_t^{i}$ as the 
time-$t$ value of overnight (O/N) rate of the collateral currency used by the party $i$.
It is not equal to the risk-free rate $r_t$, in general, which is necessary
to make the system consistent with the cross currency market~\footnote{See Ref.~\cite{collateral_choice} for details.}.
We denote the recovery rate of the party $i$ by $R_t^i\in[0,1]$.
We assume that all the processes except default times, such as $\{c^i, r, \delta^i, R^i\}$ are 
adapted to the background filtration $\mathbb{G}$.} As for the details of exposure to the counter party 
and the recovery scheme, 
see \cite{asymmetric_collateral}.

\subsection{CDS Pricing}
We denote the CDS reference name by party-$0$, the investor by party-$1$,
and the counter party by party-$2$, respectively. Let us define $\tau=\tau^0\wedge\tau^1\wedge\tau^2$ and 
its corresponding indicator process, $H_t=\bold{1}_{\{\tau\leq t\}}$.
We assume that the investor is a protection buyer and party-$2$ is a seller.
Under this setup, the CDS price from the view point of the investor can be written as
\bea
S_t&=&\beta_tE^Q\left[\int_{]t,T]}\beta_u^{-1}\bold{1}_{\{\tau>u\}}\Bigl(dD_u+q(u,S_u)S_u du\Bigr)\right. \nonumber\\
&+&\left.\left.\int_{]t,T]}\beta_u^{-1}\bold{1}_{\{\tau\geq u\}}
\Bigl(Z_u^0 dH_u^0+Z^1(u,S_{u-})dH_u^1+Z^2(u,S_{u-})dH_u^2\Bigr)\right|\calf_t\right]~,
\label{cds}
\eea
where $D$ denotes the cumulative dividend process representing the premium payment for the CDS,
and $\beta_t=\exp\left(\int_0^t r_s ds\right)$ denotes the money-market account with 
the risk-free interest rate. 
Other variables are defined as follows (See also \cite{asymmetric_collateral} for details.):
\bea
q(t,v)&=&\delta_t^1 y_t^1\bold{1}_{\{v<0\}}+\delta_t^2 y_t^2 \bold{1}_{\{v\geq 0\}}\nonumber\\
Z^0_t&=&(1-R_t^0)\nonumber\\
Z^1(t,v)&=&\Bigl(1-(1-R_t^1)(1-\delta_t^1)^+\Bigr)v\bold{1}_{\{v<0\}}+
\Bigl(1+(1-R_t^1)(\delta_t^2-1)^+\Bigr) v\bold{1}_{\{v\geq 0\}}\nonumber\\
Z^2(t,v)&=&\Bigl(1-(1-R_t^2)(1-\delta_t^2)^+\Bigr)v\bold{1}_{\{v\geq 0\}}+
\Bigl(1+(1-R_t^2)(\delta_t^1-1)^+\Bigr)v \bold{1}_{\{v<0\}}~.\nonumber
\eea
Here, $y_t^i=r_t^i-c_t^i$ is the difference of the risk-free and collateral rates relevant for the 
collateral currency chosen by the party-$i$ at the time $t$, which 
represents the instantaneous return of the posted collateral.
Thus the term $q(t,v)$ summarizes the return (or cost) of collateral from the view point of the investor.
$Z^i$ represents the default payoff when party-$i$ defaults first among the set $\{0,1,2\}$.

Although we can follow the same procedures used in the previous work~\cite{asymmetric_collateral} 
based on the arguments of Duffie \& Huang (1996)~\cite{Duffie}, we need a careful 
treatment to avoid the jump in the value process at the time of counter party 
default~\footnote{See the remark just after the proposition~$1$ in \cite{asymmetric_collateral}.}.
In this work, we 
apply
the measure change technique introduced by 
Sch\"onbucher~\cite{Schonbucher} and used by Collin-Dufresne et.al. (2004)~\cite{collin},
which leads to the following proposition in a clearcut way.

\begin{proposition}
Under the assumptions given in \ref{setup} and appropriate integrability conditions, 
the pre-default value $V_t$ corresponding to the 
CDS contract specified in Eq.~(\ref{cds}) is given by
\be
V_t=E^{Q^\prime}\left[\left.\int_{]t,T]}\exp\left(-\int_t^s \bigl(r_u-\mu(u,V_u)+h_u^0\bigr)du
\right)\Bigl(dD_s+Z_s^0 h_s^0 ds\Bigr)\right|\calf_t^\prime\right]~,
\ee
where
\bea
\mu(u,v)&=&\Bigl(y_t^1\delta_t^1-(1-R_t^1)(1-\delta_t^1)^+h_t^1+(1-R_t^2)(\delta_t^1-1)^+h_t^2\Bigr)\bold{1}_{\{v<0\}}\nonumber\\
&&+\Bigl(y_t^2\delta_t^2-(1-R_t^2)(1-\delta_t^2)^+h_t^2+(1-R_t^1)(\delta_t^2-1)^+h_t^1\Bigr)\bold{1}_{\{v\geq 0\}}~,
\eea
and it satisfies $S_t=\bold{1}_{\{\tau>t\}}V_t$ for all $t\geq 0$.
Here, the "survival measure" $Q^\prime$ is defined by 
\be
\left.\frac{dQ^\prime}{dQ}\right|_{\calf_t}=\prod_{i=0}^2\bold{1}_{\{\tau^i>t\}}\exp
\left(\int_0^t \sum_{i=0}^2 h_s^i ds\right)
\label{survival_m}
\ee
and the filtration $\mathbb{F}^\prime=(\calf_t^\prime)_{t\geq 0}$ denotes the augmentation of $\mathbb{F}$ under $Q^\prime$.
\end{proposition}
{\it Proof}: Using the Doob-Meyer decomposition, one obtains
\bea
S_t&=&\beta_t E^Q\left[\int_{]t,T]}\beta_u^{-1}\bold{1}_{\{\tau>u\}}\Bigl(dD_u+q(u,S_u)S_u du\Bigr)\right.\nonumber\\
&&+\left.\left.\int_t^T \beta_u^{-1}\bold{1}_{\{\tau>u\}}\Bigl(
Z_u^0 h_u^0 +Z^1(u,S_u)h_u^1 + Z_u^2(u,S_u)h_u^2\Bigr)du\right|\calf_t\right]~.
\eea
Let us define 
\be
\eta_t=\left.\frac{dQ^\prime}{dQ}\right|_{\calf_t}=\bold{1}_{\{\tau>t\}}\Lambda_t~,
\ee
where we have used
\be
\Lambda_t= \exp\left(\int_0^t \tilde{h}_s ds\right)~.
\ee
and $\tilde{h}_s=\sum_{i=0}^2 h_s^i$. Then, we can proceed as
\bea
S_t&=&\beta_t E^Q\left[\int_{]t,T]} \left(\beta_u\Lambda_u\right)^{-1}\eta_u
\Bigl(dD_u+q(u,S_u)S_u du\Bigr)\right. \nonumber\\
&&+\left.\left.\int_t^T \left(\beta_u \Lambda_u\right)^{-1}\eta_u \Bigl(Z_u^0 h_u^0
+Z^1(u,S_u)h_u^1 +Z^2(u,S_u)h_u^2\Bigr)du\right|\calf_t\right]\nonumber\\
&&\hspace{-12mm}=\bold{1}_{\{\tau>t\}}E^{Q^\prime}\left[
\left.\int_{]t,T]}\frac{\beta_t\Lambda_t}{\beta_u\Lambda_u}
\left\{ dD_u+\Bigl(q(u,S_u)S_u + Z_u^0 h_u^0 +Z^1(u,S_u)h_u^1 +Z^2(u,S_u)h_u^2 \Bigr) du\right\}
\right|\calf_t^\prime\right]~.\nonumber
\eea
Thus, on the set $\{\tau>t\}$, we can write
\be
V_t=E^{Q^\prime}\left[
\left.\int_{]t,T]}\frac{\beta_t\Lambda_t}{\beta_u\Lambda_u}
\left\{ dD_u+\Bigl(q(u,V_u)V_u + Z_u^0 h_u^0 +Z^1(u,V_u)h_u^1 +Z^2(u,V_u)h_u^2 \Bigr) du\right\}
\right|\calf_t^\prime\right]~.\nonumber
\ee
Simple algebra gives us
\be
V_t=E^{Q^\prime}\left[\left.\int_{]t,T]}e^{-\int_t^s (r_u+\tilde{h}_u)du}
\left\{dD_s+Z_s^0h_s^0ds+\Bigl(\tilde{h}_s-h_s^0+\mu(s,V_s)\Bigr)V_s ds\right\}\right|\calf_t^\prime\right]~.\nonumber
\ee
Integrating the linear terms gives us the desired result.~$\blacksquare$
\\
\\
Now, the method used in \cite{asymmetric_collateral} gives us the CDS valuation formula under various situations 
of collateralization. Let us summarize some of the useful results here:
\begin{corollary}
Assume the perfect and symmetric collateralization with domestic currency $(\delta^1=\delta^2=1, y^1=y^2=y)$. Then, we have
\be
V_t=E^{Q^\prime}\left[\left.
\int_{]t,T]}\exp\left(-\int_t^s (c_u+h_u^0) du\right)\Bigl(dD_s+Z_s^0 h_s^0 ds\Bigr)\right|\calf_t^\prime\right]~
\ee
as the pre-default value of the CDS given in Eq~(\ref{cds}).
\end{corollary}

\begin{corollary}
Assume the perfect and symmetric collateralization but the collateral currency $(j)$~\footnote{We use
"$()$" to denote the currency type instead of counter party.} is different from the 
evaluation currency $(i)$. In this case, we have
\be
V_t=E^{Q_{(i)}^{\prime}}\left[\left.
\int_{]t,T]}\exp\left(-\int_t^s (c_u^{(i)}+y_u^{(i,j)}+h_u^0) du\right)\Bigl(dD_s+Z_s^0 h_s^0 ds\Bigr)\right|\calf_t^\prime\right]~
\ee
where $y^{(i,j)}=y^{(i)}-y^{(j)}$. $Q_{(i)}$ and the associated $Q^\prime_{(i)}$ denote the measure 
related to the money-market account of currency $(i)$.
\end{corollary}
Notice that, as we have demonstrated in the work~\cite{collateral_choice}, the term structure of $y^{(i,j)}$, 
which is the spread of collateral cost between the two currencies,
can easily be bootstrapped from the cross currency swap information in the market.
\\

It is also straightforward to derive the leading order approximation 
under the imperfect collateralization, or $\delta^i\neq 1$, using the Gateaux derivative.
See \cite{asymmetric_collateral} for the details.
\begin{corollary}
Assume symmetric collateralization with domestic currency, but consider the 
imperfect collateralization $\delta_t^i \neq 1$.
In this case, in the leading order approximation, we have
\be
V_t\simeq \overline{V}_t+(\nabla V)_t^1 +(\nabla V)^2_t
\ee
where
\be
\overline{V}_t=E^{Q^\prime}\left[\left.\int_{]t,T]}\exp\left(-\int_t^s(c_u+h_u^0)du\right)
\Bigl(dD_s+Z_s^0h_s^0 ds\Bigr)\right|\calf_t^\prime\right]~
\ee
and
\bea
(\nabla V)^1_t&=&E^{Q^\prime}\left[\left.\int_t^T e^{-\int_t^s (c_u+h_u^0)du}
y_s \Bigl\{(1-\delta_s^1)[-\overline{V}_s]^+-(1-\delta_s^2)[\overline{V}_s]^+\Bigr\}ds\right|
\calf_t^\prime\right]\\
(\nabla V)^2_t&=&E^{Q^\prime}\left[\left.\int_t^T
e^{-\int_t^s (c_u+h_u^0)du}(1-R_s^1)h_s^1\Bigl\{(1-\delta^1_s)^+[-\overline{V}_s]^++(\delta_s^2-1)^+
[\overline{V}_s]^+\Bigr\}ds\right|\calf_t^\prime\right]\nonumber\\
&-&E^{Q^\prime}\left[\left.
\int_t^T e^{-\int_t^s (c_u+h_u^0)du}(1-R_s^2)h_s^2\Bigl\{
(1-\delta_s^2)^+[\overline{V}_s]^++(\delta_s^1-1)^+[-\overline{V}_s]^+\Bigr\}ds\right|\calf_t^\prime\right]~.\nonumber\\
\eea
\end{corollary}
In the previous work~\cite{asymmetric_collateral}, we have interpreted
$(\nabla V)^1_t$ as CCA (collateral cost adjustment) and $(\nabla V)^2_t$ as CVA.
Although it still seems a reasonable interpretation, we need to be more careful about the 
interpretation of $\overline{V}_t$, the CDS value under the perfect collateralization.
In fact, in the reminder of the paper, we will concentrate on this value 
and we will soon realize that the proper understanding of this simplest situation is 
quite non-trivial and critical for the risk management of CDS trades.

\section{Financial Implications}
Let us concentrate on the simplest case, where the symmetric and perfect collateralization is done 
continuously with domestic currency.  As we have seen, the pre-default value of the CDS
is given by 
\be
V_t=E^{Q^\prime}\left[\left.
\int_{]t,T]}\exp\left(-\int_t^s (c_u+h_u^0) du\right)\Bigl(dD_s+Z_s^0 h_s^0 ds\Bigr)\right|\calf_t^\prime\right]~
\label{perfect_collateral}
\ee
and one can easily confirm that the hazard rates of the investor $1$ as well as the counter party $2$ are absent 
from the pricing formula. Naively, it looks as if we succeed to recover the risk-free situation 
by the stringent collateral management. However, we will just see that it is quite misleading and dangerous
to treat the result of Eq.~(\ref{perfect_collateral}) as the usual risk-free pricing formula.

The key point resides in the new measure $Q^\prime$ and the filtration $\mathbb{F}^\prime$.
As was emphasized in the works~\cite{Schonbucher, collin}, the transformation in Eq.~(\ref{survival_m})
puts zero weight on the events where the parties $\{0,1,2\}$ default.
It can be easily checked as follows: By construction, we know that
\be
M_t=H_t-\int_0^t(1-H_s)\tilde{h}_s ds 
\ee
is a $(Q,\mathbb{F})$-martingale. Then, Maruyama-Girsanov's theorem implies that
\be
M^\prime_t=M_t-\int_0^t\frac{d\langle M,\eta\rangle_s}{\eta_{s-}}
\ee
should be a $(Q^\prime, \mathbb{F}^\prime)$-martingale, 
where $\langle \cdot,\cdot\rangle$ denotes the (conditional or predictable) quadratic covariation.
Now, one can easily check that
\bea
M^\prime_t&=&M_t+\int_0^t (1-H_s)\tilde{h}_s ds\nonumber\\
 &=&H_t~
\eea
and thus $H_t=\bold{1}_{\{\tau\leq t\}}$ itself becomes a $(Q^\prime, \mathbb{F}^\prime)$-martingale.
In other words, under the new measure, the parties $\{0,1,2\}$ do not default almost surely.

Let us consider the financial implications of this fact.
By our construction of filtration, $(Q,\mathbb{F})$ hazard rate process of party $i$ can be 
written in the following form in general:
\be
h_t^i=\sum_{\{ D\in\Pi;~i\not\in D\}}\prod_{j\in D} \bold{1}_{\{\tau^j\leq t\}}\prod_{k\in\calc\backslash D}
\bold{1}_{\{\tau^k>t\}}h_D^i(t)
\label{F-intensity}
\ee
where $\Pi$ denotes the set of all the subgroups of $\calc=\{0,1,2,\cdots,n\}$ including also the empty set.
Let us define $\mathbb{H}^D=\prod_{j\in D}\vee \mathbb{H}^k$, then $h^i_D(t)$ is adapted to the 
filtration $\mathbb{G}\vee \mathbb{H}^D$, and hence includes the information of default times
of the parties included in the set $D$ although they are not explicitly shown in the formula.
Importantly, it is not the hazard rate (or default intensity) under the new measure $(Q^\prime, \mathbb{F}^\prime)$.

In the new measure, the parties included in $\cals=\{0,1,2\}$ are almost surely alive for all $t\geq 0$,
and hence $h_t^i$ given in Eq.~(\ref{F-intensity}) is equivalent to $h_t^{\prime i}$ given below:
\be
h_t^{\prime i}=\sum_{\{D\in \Pi^\prime;~i\not\in D\}}\prod_{j\in D}\bold{1}_{\{\tau^j\leq t\}}
\prod_{k\in \calc^\prime\backslash D} \bold{1}_{\{\tau^k>t\}}h_D^i(t)
\label{hprime}
\ee
where $\calc^\prime=\calc\backslash \cals$, and $\Pi^\prime$ denotes 
the set of all the subgroups of $\calc^\prime$ and the empty set.
Thus, the pre-default value of the CDS can also be written as
\be
V_t=E^{Q^\prime}\left[\left.
\int_{]t,T]}\exp\left(-\int_t^s (c_u+h_u^{\prime 0}) du\right)\Bigl(dD_s+Z_s^0 h_s^{\prime 0} ds\Bigr)\right|\calf_t^\prime\right]~.
\label{cds_newm}
\ee
Now, let us consider the difference from the situation where the same CDS is being traded 
between the two completely default risk free parties. 
In this case, the pre-default value of the CDS is given by
\be
V_t^{\rm rf}=E^{Q}\left[\left.
\int_{]t,T]}\exp\left(-\int_t^s (c_u+h_u^0) du\right)\Bigl(dD_s+Z_s^0 h_s^0 ds\Bigr)\right|\calf_t \right]~.
\ee
Here, we have assumed that the collateralization is still being carried out.
It is now clear that the most important quantity to understand is 
\be
V_t^{\rm rf}-V_t~.
\ee
Considering $t=0$ allows us to have a better image. In this case, $V_0^{\rm rf}$ can be
simply obtained by the knowledge of marginal default distribution of the reference entity.
On the other hand, $V_0$ does not contain the contribution from the scenarios 
where the parties included in the set $\cals$ default in the future, and hence the value of protection 
from the CDS becomes smaller than the former one.

In order to understand this fact, it is instructive to consider the predictable nature of 
collateralization. Even under the perfect collateralization, the party can only recover the contract 
value just before the default of the counter party.  In the case of non-credit sensitive 
products, such as the standard interest rate swaps e.t.c., it does not cause any 
problem and we can consider the valuation just as if we are in risk-free world 
once we take the cost of collateral into account appropriately.
However, in the case of credit derivatives, the contract value is expected to change
discretely following the jump of the relevant hazard rate because of the direct feedback 
or contagious effect from the counter party default.
Although the marginal default intensity contains the contribution from all the scenarios,
the investor cannot receive this value in reality.
In addition, the investor cannot receive any return caused by his default and all
the following events after it.
Thus, the quantity $V^{\rm rf}-V$ can be interpreted as 
the net discounted value that should be obtained by the investor if he and his counter party
were default free. 

It is easy to imagine that the quantity is determined 
by the size of the price jump caused by contagious (or feedback) effects from the defaults of the two parties
and hence the default dependence among the relevant names.
Although it is difficult to obtain the difference in a generic situation, 
we will demonstrate its significance in a simple but important setup in the following 
sections.
\\

{\it Remark}:~In the market, the claim to the defaulted entity is to be made
just {\it after} the default event. Thus, some additional value may be recovered on top of the collateral amount,
because the claim is based on the market value {\it after} the default. 
This additional claim can be interpreted as
the substitution cost, but the proper theoretical handling is difficult since it does depend on 
the choice of the next counter party and the exact timing of execution.
Although ISDA document allows to include the cost of actual replacements,
it is now clear that this is a huge burden of complicated litigious works on the administrator of the defaulted company.
Since it is very difficult to prove the cost of replacement claimed by the creditors, such as the bid/offer spreads
at the time of executions, is actually fair. It also crucially depends on 
whether the set of trades were unwound in a net basis or one-by-one, which should have 
a significant impact on the modeling of recovery scheme. 

According to the recent article~\cite{risk_article}, the settlement of claims against Lehman Brothers 
has completed only 11\% up to now.
It seems that we need a clear rule to guarantee the fairness among the 
creditors and  to speed up the settlement process.
Once the detailed procedures and the rights of a creditor are
finely defined, more accurate recovery handling will be possible.

\section{Special Cases}
\label{specialcase}
We first list up the two special cases which provide us better 
understanding. Especially, the second example can be applied to the back-to-back trade,
which is the most relevant situation for the CCPs.
\subsection{3-party Case}
The simplest situation to calculate the collateralized CDS is the 
case where there are only three relevant names, $\calc=\{0,1,2\}$, 
which are the reference entity, the investor and the counter party, respectively.
In this case, since $\cals=\calc$, the set $\Pi^\prime$ contains only the 
empty set $\{\emptyset \}$.
Thus, under the survival measure $(Q^\prime,\mathbb{F}^\prime)$, we have 
\be
h^{\prime i}_t = h^i_{\{\emptyset\}}(t)
\ee
and particularly,
\be
h^{\prime 0}_t=h^0_{\{\emptyset\}}(t)~.
\ee
Since we know that $h_{\{\emptyset\}}^i$ is adapted to the background filtration $\mathbb{G}$,
the evaluation of CDS value is quite straightforward in this case.

\subsection{4-party Case}
Now, let us add one more party and consider the 4-party case, $\calc=\{0,1,2,3\}$.
We first consider the trade of a CDS between the investor (party-$1$) and 
the counter party (party-$2$). The reference entity is party-$0$. 
In this case, we have $\calc^\prime =\{3\}$.
Therefore, the relevant processes under the survival measure are 
\bea
h_t^{\prime 0}&=&\bold{1}_{\{\tau^3>t\}}h^0_{\{\emptyset\}}(t)+\bold{1}_{\{\tau^3\leq t\}}
h_{\{3\}}^0(t,\tau^3)\\
h_t^{\prime 3}&=&\bold{1}_{\{\tau^3>t\}}h^3_{\{\emptyset\}}(t) ~.
\eea
Here, we have made the dependence on the default time in $h^0_{\{3\}}$ explicitly.
Since the feedback effect only appears through the party $3$ in the survival measure, 
we can proceed in similar fashion done in Ref.~\cite{collin}.

In this case, the perfectly collateralized CDS pre-default value turns out to be 
\bea
V_t&=&\bold{1}_{\{\tau^3\leq t\}}E^{Q^\prime}
\left[\left. \int_{]t,T]}e^{-\int_t^s \bigl(c_u+h^0_{\{3\}}(u,\tau^3)\bigr)du}
\Bigl(dD_s+Z_s^0 h_{\{3\}}^0(s,\tau^3)ds\Bigr)\right|\calf_t^\prime\right]\nonumber\\
&&\hspace{-22mm}+\bold{1}_{\{\tau^3>t\}}\left\{
E^{Q^\prime}\left[\left.\int_{]t,T]}e^{-\int_t^s \bigl(c_u+h_{\{\emptyset\}}^0(u)\bigr)du}
\left(e^{-\int_t^s h^3_{\{\emptyset\}}(u)du}\right)\Bigl(dD_s+Z_s^0 h^0_{\{\emptyset\}}(s)ds\Bigr)
\right|\calf_t^\prime\right]\right.\nonumber \\
&&\hspace{-22mm}+E^{Q^\prime}\left[\left.\int_{]t,T]}e^{-\int_t^s c_u du}
\left( \int_t^s e^{-\left(\int_t^v h^0_{\{\emptyset\}}(u)+\int_v^s h^0_{\{3\}}(u,v)\right)du}
\left[ e^{-\int_t^v h^3_{\{\emptyset\}}(x)dx} h^3_{\{\emptyset\}}(v)\right] dv\right)
dD_s\right|\calf_t^\prime\right]\nonumber\\
&&\hspace{-22mm}\left.+E^{Q^\prime}\left[\left.\int_{]t,T]} e^{-\int_t^s c_udu}
\left(\int_t^s  e^{-\left(\int_t^v h^0_{\{\emptyset\}}(u)+\int_v^s h^0_{\{3\}}(u,v)\right)du}
\left[ e^{-\int_t^v h^3_{\{\emptyset\}}(x)dx} h^3_{\{\emptyset\}}(v)\right]
h^0_{\{3\}}(s,v)dv\right)Z_s^0 ds\right|\calf_t^\prime\right]\right\}~.\nonumber\\
\eea
Here, the important point is not the possible contagious effect from the party-$3$,
but rather the {\it lack} of the contagious effects from the other names included in the 
set $\cals$, which are included in the marginal default intensity of the reference entity.

Now, because of the symmetry, if the investor enters a back-to-back trade with the 
counter party $3$, the pre-default value of this offsetting contract is given as follows:
\bea
V_t^{B2B}&=&-\bold{1}_{\{\tau^2\leq t\}}E^{Q^{\prime\prime}}
\left[\left. \int_{]t,T]}e^{-\int_t^s \bigl(c_u+h^0_{\{2\}}(u,\tau^2)\bigr)du}
\Bigl(dD_s+Z_s^0 h_{\{2\}}^0(s,\tau^2)ds\Bigr)\right|\calf_t^{\prime\prime}\right]\nonumber\\
&&\hspace{-22mm}-\bold{1}_{\{\tau^2>t\}}\left\{
E^{Q^{\prime\prime}}\left[\left.\int_{]t,T]}e^{-\int_t^s \bigl(c_u+h_{\{\emptyset\}}^0(u)\bigr)du}
\left(e^{-\int_t^s h^2_{\{\emptyset\}}(u)du}\right)\Bigl(dD_s+Z_s^0 h^0_{\{\emptyset\}}(s)ds\Bigr)
\right|\calf_t^{\prime\prime}\right]\right.\nonumber \\
&&\hspace{-22mm}+E^{Q^{\prime\prime}}\left[\left.\int_{]t,T]}e^{-\int_t^s c_u du}
\left( \int_t^s e^{-\left(\int_t^v h^0_{\{\emptyset\}}(u)+\int_v^s h^0_{\{2\}}(u,v)\right)du}
\left[ e^{-\int_t^v h^2_{\{\emptyset\}}(x)dx} h^2_{\{\emptyset\}}(v)\right] dv\right)
dD_s\right|\calf_t^{\prime\prime}\right]\nonumber\\
&&\hspace{-22mm}\left.+E^{Q^{\prime\prime}}\left[\left.\int_{]t,T]} e^{-\int_t^s c_udu}
\left(\int_t^s  e^{-\left(\int_t^v h^0_{\{\emptyset\}}(u)+\int_v^s h^0_{\{2\}}(u,v)\right)du}
\left[ e^{-\int_t^v h^2_{\{\emptyset\}}(x)dx} h^2_{\{\emptyset\}}(v)\right]
h^0_{\{2\}}(s,v)dv\right)Z_s^0 ds\right|\calf_t^{\prime\prime}\right]\right\}~.\nonumber\\
\eea
Here, $Q^{\prime\prime}$ is defined according to the new survival set $\cals^{B2B}=\{0,1,3\}$ by
\be
\left.\frac{dQ^{\prime\prime}}{dQ}\right|_{\calf_t}=\prod_{i\in \cals^{B2B}}\bold{1}_{\{\tau^i>t\}}\exp
\left(\int_0^t \sum_{i\in \cals^{B2B}} h_s^i ds\right)
\ee
and the filtration $\mathbb{F}^{\prime\prime}=(\calf_t^{\prime\prime})_{t\geq 0}$ denotes the augmentation 
of $\mathbb{F}$ under $Q^{\prime\prime}$.

Now let us consider $V_0+V_0^{B2B}$.  It is easy to check that it is not zero in general
and does depend on the default intensities of party-$2$ and -$3$, and also their contagious 
effects to the reference entity. Suppose that the investor is a CCP just entered into the 
back-to-back trade with the party-$2$ and -$3$ who have the same 
marginal default intensities. Even under the perfect collateralization, 
if the CCP applies the same CDS price (or premium) to the two parties, it has, in general, the mark-to-market loss or profit
even at the inception of the contract. For example, consider the case where the protection seller (party-2) has very high
default dependence with the reference entity, while the buyer (party-3) of the  protection from the CCP has 
smaller one. 
In this case, the $h^0$ in $(Q^\prime,\mathbb{F}^\prime)$ should be smaller than that of 
$(Q^{\prime\prime},\mathbb{F}^{\prime\prime})$.
If the CCP uses the same CDS premium to the two parties, CCP should properly recognizes the 
loss, which stems from the difference of the contagion size from the default of the two parties.
Since the party $2$ has high default dependence, the short protection position of the CCP 
with the party $3$ suffers bigger loss at the time of the default of the party $2$.

\section{Examples using a Copula}
In order to separate the marginal intensity and default dependence, we will adopt here
the copula framework. After explaining the general setup, we will apply Clayton copula
to demonstrate quantitative impact.
\subsection{Framework}
Suppose that we are given a non-negative process $\lambda^i$ adapted to the 
background filtration $\mathbb{G}$ for each party $i\in \calc=\{0,1,2,\cdots, n\}$.
Suppose also that there exists a uniformly distributed random variable $U^i\in[0,1]$.
We assume that, under $(Q,\calf_0)$, the $(n+1)$-dimensional random vector
\be
\vec{U}=(U^0,U^1,\cdots, U^n) 
\ee
is distributed according to the $(n+1)$-dimensional copula
\be
C(\vec{u})~.
\ee
We further assume that $\vec{U}$ is independent from $\calg_{\infty}$
and also that copula function $C$ is $(n+1)$-times continuously differentiable.

Now, let us define the default time $\tau^i$ as 
\be
\tau^i = \inf\Bigl\{t; ~e^{-\int_0^t \lambda_s^i ds}\leq U^i\Bigr\}~.
\ee
Then, given the information $\calg_{\infty}$, one obtains the joint default distribution as
\be
Q\left.\Bigl(\tau^0>T^0, \tau^1>T^1, \cdots, \tau^n>T^n\right|\calg_{\infty}\Bigr)=C\bigl(\vec{\gamma}(\vec{T})\bigr)
\ee
where we have used the notation  of
\be
C\bigl(\vec{\gamma}(\vec{T})\bigr) = C\bigl(\gamma^0(T^0), \cdots, \gamma^n(T^n)\bigr)
\ee
and 
\be
\gamma^i(T)=\exp\left(-\int_0^T \lambda_s^i ds\right)~.
\ee

Following the well known procedures~\footnote{See, 
the works\cite{copula1,copula2}, for example.},  one obtains
\bea
Q^i(t,T)&=&Q\left(\left. \bold{1}_{\{\tau^i>T^i\}} \right|\calf_t\right)\nonumber\\
&&\hspace{-22mm}=\sum_{\{D\in \Pi;~i\not\in D\}}\prod_{j\in D}\bold{1}_{\{\tau^j\leq t\}}
\prod_{k\in \calc\backslash D}\bold{1}_{\{\tau^k>t\}}
\left.\frac{E^Q\left[\left.\vec{\partial}_{D}C\bigl(\gamma^i(T^i), \vec{\gamma}^{\calc \backslash \{D,i\}}(t),
\vec{\gamma}^D(\vec{t}^{D})\bigr)
\right|\calg_t\right]}{\vec{\partial}_D C\bigl(\vec{\gamma}^{\calc\backslash D}(t),\vec{\gamma}^D(\vec{t}^{D})\bigr)}\right|_{\vec{t}^D=\vec{\tau}^D}~.
\nonumber\\
\label{zcb_copula}
\eea
Here, we have defined $\Pi$ as the set containing all the subgroups of $\calc$ with the empty set, and 
\be
\vec{\partial}_D=\prod_{i\in D}\frac{\partial}{\partial u^i}~.
\ee
$\vec{\gamma}^D(\vec{\tau^D})$ is the set of $\gamma^i(\tau^i)$ for all $i\in D$, and similarly for $\vec{\gamma}^{\calc\backslash D}$.
In the expression of Eq.~(\ref{zcb_copula}), we have not properly ordered the arguments of the copula function
just for simplicity, which should be understood in the appropriate way.

$(Q,\mathbb{F})$ hazard rate of party-$i$ is calculated as
\bea
h^i_t&=&\left.-\frac{\partial}{\partial T}\ln Q^i(t,T)\right|_{T=t}\nonumber\\
&&\hspace{-20mm}=\lambda^i_t\gamma^i(t)\sum_{\{D\in \Pi;~i\not\in D\}}\prod_{j\in D}\bold{1}_{\{\tau^j\leq t\}}
\prod_{k\in \calc\backslash D}\bold{1}_{\{\tau^k>t\}}
\frac{\partial_i\vec{\partial}_D C\bigl(\vec{\gamma}^{\calc\backslash D}(t),\vec{\gamma}^D(\vec{\tau}^D)\bigr)}{
\vec{\partial}_DC\bigl(\vec{\gamma}^{\calc\backslash D}(t),\vec{\gamma}^D(\vec{\tau}^D)\bigr)}~.
\eea
Hence, as we have done in Eq~(\ref{hprime}) for the survival set $\cals=\{0,1,2\}$,
$h^i_t$ can be replaced as follows under the new measure $(Q^\prime, \mathbb{F}^\prime)$:
\be
h^{\prime i}_t=\lambda^i_t\gamma^i(t)\sum_{\{D\in \Pi^\prime;~i\not\in D\}}\prod_{j\in D}\bold{1}_{\{\tau^j\leq t\}}
\prod_{k\in \calc^\prime\backslash D}\bold{1}_{\{\tau^k>t\}}
\frac{\partial_i\vec{\partial}_D C\bigl(\vec{\gamma}^{\calc\backslash D}(t),\vec{\gamma}^D(\vec{\tau}^D)\bigr)}{
\vec{\partial}_DC\bigl(\vec{\gamma}^{\calc\backslash D}(t),\vec{\gamma}^D(\vec{\tau}^D)\bigr)}~,
\ee
where $\calc^\prime=\calc\backslash \cals$ and $\Pi^\prime$ is the set of all the subgroups of $\calc^\prime$
with an additional empty set.
One can observe that both $h^{i}$ and $h^{\prime i}$ are equal to the marginal intensity $\lambda^i$
if there is no default dependence (or in the case of the product copula).
\\

Now, let us apply the copula framework to the two special cases given in Sec.~\ref{specialcase}.\\
\large (3-party Case):~\normalsize 
In the 3-party case with $\calc=\{0,1,2\}$, we have
\be
h^{\prime 0}_t=\lambda^0_t\gamma^0(t)\frac{\partial_0 C\bigl(\vec{\gamma}(t)\bigr)}{C\bigl(\vec{\gamma}(t)\bigr)}~,
\ee
where $\vec{\gamma}(t)=(\gamma^0(t),\gamma^1(t),\gamma^2(t))$~.
\\
\\
\large (4-party Case):~\normalsize
In the 4-party case with $\calc=\{0,1,2,3\}$, we have
\bea
h^{\prime 0}_t&=&\bold{1}_{\{\tau^3>t\}}
\frac{\partial_0 C\bigl(\vec{\gamma}(t)\bigr)}{C\bigl(\vec{\gamma}(t)\bigr)}
+\bold{1}_{\{\tau^3\leq t\}}\frac{\partial_0\partial_3 C\bigl(\vec{\gamma}^{\calc \backslash \{3\}}(t),\gamma^3(\tau^3)\bigr)}{
\partial_3 C\bigl(\vec{\gamma}^{\calc\backslash \{3\}}(t),\gamma^3(\tau^3)\bigr)}\\
h^{\prime 3}_t&=&\bold{1}_{\{\tau^3>t\}}\frac{\partial_3C\bigl(\vec{\gamma}(t)\bigr)}{C\bigl(\vec{\gamma}(t)\bigr)}
\eea
under $(Q^\prime, \mathbb{F}^\prime)$, and 
\bea
h^{\prime\prime 0}_t&=&\bold{1}_{\{\tau^2>t\}}
\frac{\partial_0 C\bigl(\vec{\gamma}(t)\bigr)}{C\bigl(\vec{\gamma}(t)\bigr)}
+\bold{1}_{\{\tau^2\leq t\}}\frac{\partial_0\partial_2 C\bigl(\vec{\gamma}^{\calc \backslash \{2\}}(t),\gamma^2(\tau^2)\bigr)}{
\partial_2 C\bigl(\vec{\gamma}^{\calc\backslash \{2\}}(t),\gamma^2(\tau^2)\bigr)}\\
h^{\prime\prime 2}_t&=&\bold{1}_{\{\tau^2>t\}}\frac{\partial_2C\bigl(\vec{\gamma}(t)\bigr)}{C\bigl(\vec{\gamma}(t)\bigr)}
\eea
under $(Q^{\prime\prime},\mathbb{F}^{\prime\prime})$.

\subsection{Numerical Examples}
In this paper, we adopt Clayton copula just for its analytical tractability and 
easy interpretation of its parameter~\cite{copula1}.
Clayton copula belongs to Archimedean copula family, whose general form is given by
\be
C(\vec{u})=\phi^{[-1]}\left(\sum_{i=0}^n \phi(u^i)\right)~,
\ee
where the function $\phi(\cdot)$ is called the generator of the copula, and $\phi^{[-1]}(\cdot)$ is its pseudo-inverse function.
For Clayton copula, the generator function is given by
\be
\phi(u)=(u^{-\alpha}-1)/\alpha
\ee
for $\alpha>0$, and hence we have
\be
C(\vec{u})=\left(-n+\sum_{i=0}^n (u^i)^{-\alpha}\right)^{-1/\alpha}~.
\ee
For this copula, one can easily check that
\bea
\gamma^i\frac{\partial_iC(\vec{\gamma})}{C(\vec{\gamma})}&=&\left(\frac{C(\vec{\gamma})}{\gamma^i}\right)^\alpha\nonumber\\
\gamma^i\frac{\partial_i\partial_j C(\vec{\gamma})}{\partial_jC(\vec{\gamma})}&=&
(1+\alpha)\left(\frac{C(\vec{\gamma})}{\gamma^i}\right)^\alpha,
\eea
which means that the hazard rate jumps to $(1+\alpha)$ times of its value just before the default of any other party.

In the following two figures, we have shown the numerical examples of the par premium 
of the perfectly collateralized CDS under Clayton copula.
For simplicity, we have assumed continuous payment of the premium, and also assumed that 
the recovery rate $R^i=40\%$, the collateral rate $c=0.02$, and the 
marginal intensity $\lambda^i$ of each name is
constant.

\begin{figure}[htbt]
	\center{\includegraphics[width=117mm]{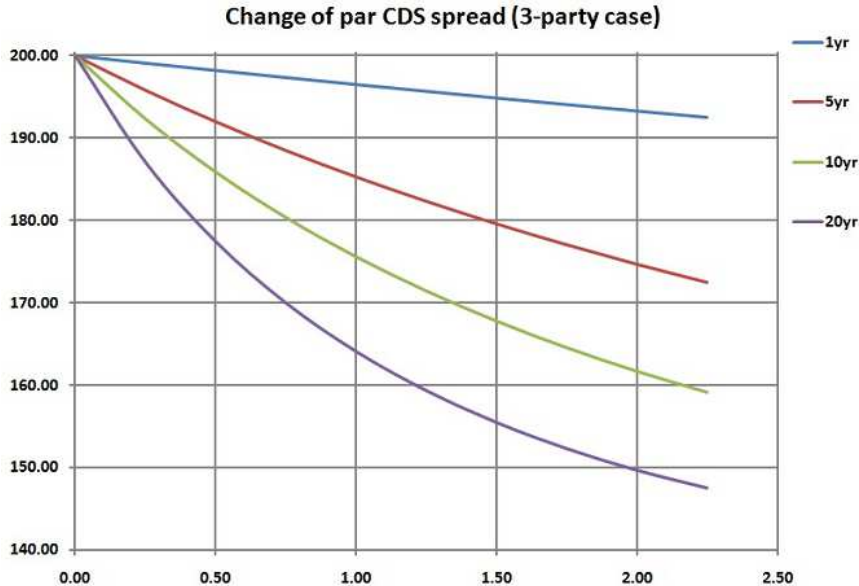}}
	\vspace{-3mm}
	\caption{Change of par CDS spread with Clayton copula parameter $\alpha$.}
	\label{3-party}
\end{figure}

In Fig.~\ref{3-party}, we have shown the results of 3-party case for the set of maturities;
1yr,~5yr,~10yr and 20yr.
Here, the effective marginal intensity of each party $\overline{\lambda}^i=(1-R^i)\lambda^i$ is 
given as follows:
\be
\bigl(\overline{\lambda}^0,\overline{\lambda}^1,\overline{\lambda}^2\bigr)=\bigl(200{\rm bp},~100{\rm bp},
~120{\rm bp}\bigr)~.\nonumber
\ee
The horizontal axis denotes the value of the copula parameter $\alpha$.
Considering the situation where major financial institutions are involved, and 
recalling the events just after the Lehman collapse, the significant jump of 
hazard rates seems possible. One can see that there is meaningful deviation of 
the par CDS premium from the marginal intensity $\overline{\lambda}^0$ even within the 
reasonable range of the jump size, or $\alpha$.

\begin{figure}[htbt]
	\center{\includegraphics[width=122mm]{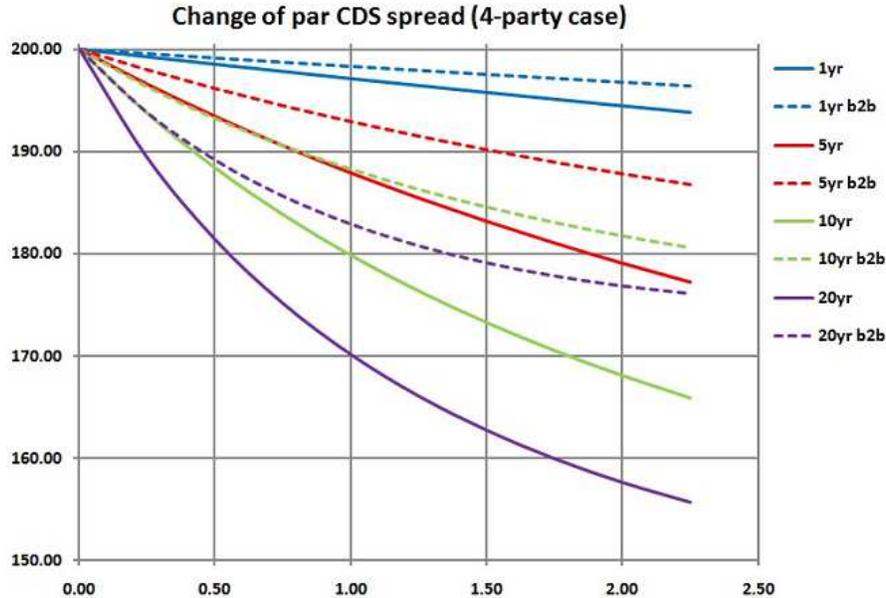}}
	\vspace{-3mm}
	\caption{Change of par CDS spread with Clayton copula parameter $\alpha$.}
	\label{4-party}
\end{figure}

If Fig.~\ref{4-party}, we have shown the corresponding results for the 4-party case.
Here, we have set the effective marginal intensities as
\be
\bigl(\overline{\lambda}^0,\overline{\lambda}^1,\overline{\lambda}^2,
\overline{\lambda}^3\bigr)=\bigl(200{\rm bp},~30{\rm bp},
~150{\rm bp},~75{\rm bp}\bigr)~.\nonumber
\ee
In this case, we have modeled the situation where the investor has very high credit quality,
which enters back-to-back trades with the two firms that have quite different 
credit worthiness. In the figure, we have used the solid lines for the 
trade with party-$2$, and the dashed lines for the offsetting trade 
with party-$3$. The result tells us that the back-to-back trades have
non-zero mark-to-market value if the investor applies the same premium.
This fact is very important for a CCP. It tells us that, even under the very stringent collateral management,
the CCP has to recognize that it is not free from the "risk". 
It is, in fact, free from the credit risk of the counter party, but still 
suffers from the contagious effects from the defaults of its counter parties.

\section{Conclusions}
In this paper, we have studied the pricing of CDS under continuous collateralization.
We have made use of the "survival measure" to avoid the "no-jump" condition 
required in the previous work~\cite{asymmetric_collateral}.
It allows us straightforward derivation of pricing formula of CDS 
under various situations of collateralization.

In the later part, we have focused on the situation where the CDS
is perfectly collateralized. We have shown that there exists
irremovable trace of the two parties in the CDS price through 
their default dependence among the relevant names.
For numerical examples, we have adopted Clayton copula to 
show the change of the par CDS premium according to the dependence parameter.
The results have shown that there exists significant deviation of the 
par premium from the marginal intensity of the reference entity 
when the default dependence is high.

We should emphasize that these numerical calculations are solely for demonstrative purpose. Because of the 
simplicity of the chosen copula, there is no way to assign different dependence 
among the relevant names. The detailed analysis with dynamic underlying processes,
and more realistic dependence structure is left for our future research.
However, we still think that our results are enough to demonstrate the importance of 
default dependence or contagious effects among the relevant names 
even under the very stringent collateral management. This fact seems particularly important
for the CCPs, where the members are usually major broker-dealers that are expected to 
have very significant impacts on all the other names just we have experienced in this crisis.


\end{document}